\documentclass[
aps,
prb,
twocolumn,
superscriptaddress,%
showpacs,%
preprintnumbers,%
byrevtex,%
floatfix%
]{revtex4-2}

\usepackage{dcolumn}
\usepackage{bm}
\usepackage{ifthen}
\usepackage[usenames]{color}
\usepackage{amssymb}
\usepackage{textcomp}
\usepackage{amsfonts}
\usepackage{amsmath}
\usepackage[dvips]{graphicx}
\usepackage{graphpap}

\begin{document}

\date{\today}
\title{Influence of (N,H)-terminated surfaces on stability, hyperfine structure, and zero-field splitting of NV centers in diamond}

\author{Wolfgang K\"orner}
\email{wolfgang.koerner@iwm.fraunhofer.de}\affiliation{Fraunhofer Institute for Mechanics of Materials IWM, W\"ohlerstra{\ss}e
11, 79108 Freiburg, Germany}

\author{Reyhaneh Ghassemizadeh}
\affiliation{Fraunhofer Institute for Mechanics of Materials IWM, W\"ohlerstra{\ss}e 
11, 79108 Freiburg, Germany}

\author{Daniel F. Urban}
\affiliation{Fraunhofer Institute for Mechanics of Materials IWM, W\"ohlerstra{\ss}e 11, 79108 Freiburg, Germany}

\author{Christian Els\"asser}
\affiliation{Fraunhofer Institute for Mechanics of Materials IWM,
W\"ohlerstra{\ss}e 11, 79108 Freiburg, Germany}%
\affiliation{University of Freiburg, Freiburg Materials Research Center (FMF), Stefan-Meier-Stra{\ss}e 21, 79104 Freiburg, Germany}

\begin{abstract}
We present a density functional theory analysis of the negatively charged nitrogen-vacancy (NV$^-$)
defect complex in diamond located in the vicinity of (111)- or (100)-oriented surfaces with mixed (N,H)-terminations. 
We assess the stability and electronic properties of the NV$^-$ center and study their dependence on the H:N ratio of the surface termination. The formation energy, the electronic density of states, the hyperfine structure and zero-field splitting parameters of an NV$^-$ center are analyzed as function of its distance and orientation to the surface. 
We find stable NV$^-$ centers with bulk-like properties at distances of at least $\sim8$ {\AA} from the surface provided that the surface termination consists of at least 25\% substitutional nitrogen atoms. The studied surface terminations have a minor effect on the ground state properties whereas the NV orientation has major effects.
Our results indicate that axial NV centers near a flat 100\% N-terminated (111) surface are the optimal choice for NV-based quantum sensing applications as they are the least influenced by the proximity of the surface. 
\end{abstract}


\maketitle
\section{Introduction}
The negatively charged nitrogen-vacancy center NV$^-$ in diamond is a point-defect complex with excellent potential for the use in spatial-atomic-resolution quantum magnetometry \cite{ba08,ma08,ac09} and in solid-state-based quantum computing.\cite{jac09}
It consists of a substitutional nitrogen atom in the diamond crystal structure with a vacant nearest-neighbor carbon site and an additional electron. 
The focus of this study is on NV$^-$ centers as functional elements in scanning magnetic-field sensors. A detailed review on the various sensing mechanisms using NV-centers can be found in Ref. \cite{lee19}
In principle, for sensing desirable to bring them as close as possible to the diamond surface in order to be able to position this atomic magnetic-field probe as close as possible to the  external magnetic field to be measured. Unfortunately, NV$^-$ centers may loose their negative charge state near surfaces with negative electron affinity (EA). A prominent example is the case of a hydrogen-terminated diamond surface where the EA is between -1 and -1.3 eV according to experiments.\cite{mai01,fu10,ro10,ha11,kaw19} 
Besides the negative EA, additional surface states may lie deeply in the band gap and interact with the electronic levels of the NV$^-$ center. 

Various theoretical studies\cite{kav14,chou17b,li19} have determined surface states for H, OH, F, and N terminations for (001), (111) and (113) surfaces. In summary, all the studied hydrogenated surfaces show deep levels below the conduction band that disturb the levels of the NV$^-$ center, whereas fluorinated and nitrogenated surfaces have only few surface states and highly positive EA. For a comprehensive overview on the surface states see Figure 1 of  Ref.\,[\onlinecite{kav14}]. A summary of the EA can be found in Table 1 of Ref.\,[\onlinecite{gal19}].

Our study deals with surfaces terminated by H atoms, N atoms, or mixtures of these two elements. Nitrogen atoms are most promising as they create the least amount of surface states while the detrimental hydrogen atoms can hardly be avoided in any experimental diamond-growth process. By considering surface terminations with different N:H ratios one can estimate to what extent H may be tolerable when aiming at the experimental realization of NV$^-$ centers very close to a surface. Therefore we want to answer the following questions: 
(i) How much hydrogen can be tolerated in a mixed (N,H)-termination of the surface in order to keep the negative charge state of the NV$^-$ center?
(ii) At what distance from the surface does a NV$^-$ center have to be situated so that its functional properties are only marginally disturbed by the proximity of the surface?

Our density functional theory (DFT) analysis is restricted to NV centers which are at most about 15 {\AA} below the surface due to computational costs. Thus we investigate the case of very shallow NV centers which has recently already been realized experimentally.  
M\"uller et al.\cite{mue14} reported single NV centers at 20-30 {\AA} and Ofori-Okai et al.\cite{ofo12} reported even implantation depths of $<$10 {\AA}. In the following the NV centers are investigated at all possible sites relative to the (001) and (111) surfaces in atomistic supercell models.
By evaluating their total energies it is clarified whether certain sites near a surface are preferred with respect to sites in the bulk crystal or not. Furthermore,  the electronic density of states (DOS), the hyperfine structure (HFS) and the zero-field splitting (ZFS) parameters are evaluated. With these results we quantify the influence of the surfaces, and we assess whether the electronic level structure of the NV$^-$ center near the surface is still intact. 

The manuscript is organized as follows: In Sec.~\ref{sec:theory} 
the details of the DFT calculations and the atomistic supercell models are described.
The results of the NV$^-$ stabilization including EA, total energy and DOS calculations are presented in Secs.~\ref{sec:results:ea},~\ref{sec:results:energy}, and \ref{sec:results:elec}, respectively. 
The results for the ZFS calculations are reported and discussed in Sec.~\ref{sec:results:zfs}, and the HFS analysis is presented in Sec.~\ref{sec:results:hyperfine}. 
Section ~\ref{sec:summary} summarizes our findings.
The Appendix gives a brief compilation of important properties of the tensors of hyperfine structure parameters $\boldsymbol{A}_{ij}^I$ and zero-field splitting parameters $\boldsymbol{D}_{ij}$.

\section{Theoretical approach}\label{sec:theory}

\subsection{Supercell models}\label{sec:supercell}

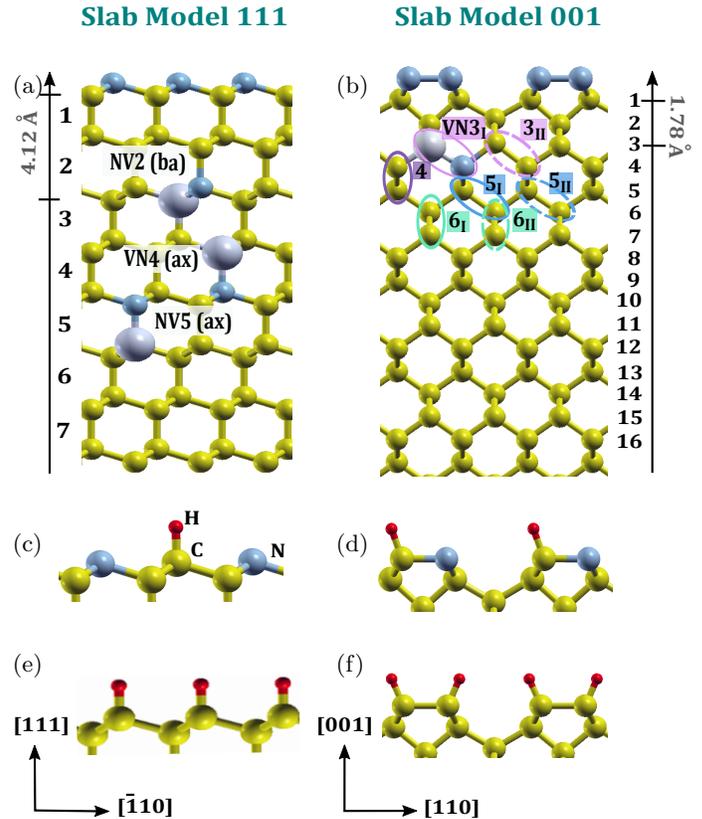
\begin{figure}{}
\begin{center}
\setlength{\unitlength}{1mm}
\begin{picture}(85,105)(0,0)
\put( 0,97){(a)}
\put(43,97){(b)}	
\put( 0,36){(c)}
\put(43,36){(d)}	
\put( 0,20){(e)}
\put(43,20){(f)}	
\put(0,0){\includegraphics[width=0.5\columnwidth,draft=false]{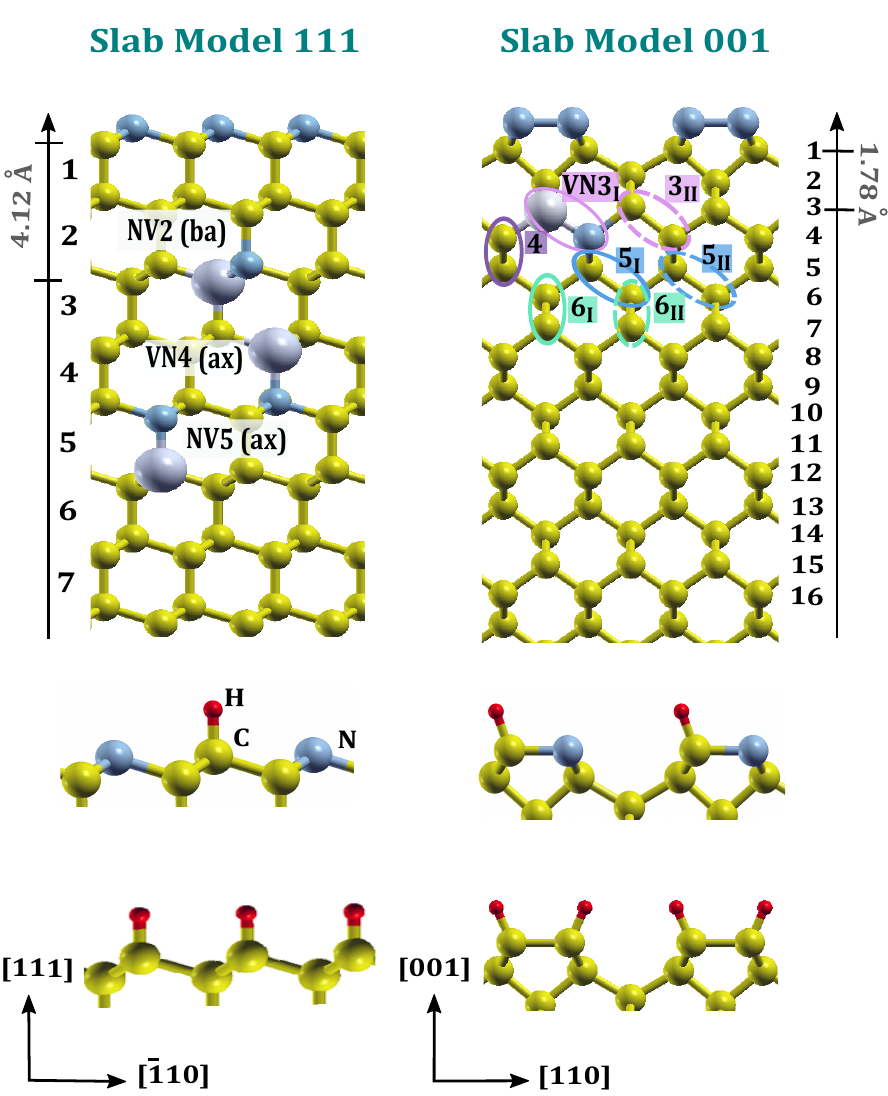}} 
\end{picture}
\end{center}
\caption{(Color online)
Atomic structure models of the (a) (111)-oriented surface and (b) (001)-oriented surface with 
100\% N termination. The terminating atomic structures for a 50\% mixed ratio of H and N as 
well as for the 100\% H termination are illustrated in (c)--(f). Yellow spheres represent  carbon atoms, small red spheres hydrogen atoms, and  blue spheres nitrogen atoms. The position of a carbon vacancy is indicated by a large gray sphere. In order to indicate the positions of the NV centers relative to the surface, the single layers for (001) and the double layers for (111) are enumerated. 
For the (001) surface the seven different (symmetry inequivalent) positions of NV centers are indicated, which differ in their relative positions with respect to the N atoms on the surface.    
For the (111) surface we plot the example of an axially oriented NV center in the fifth double layer, denoted NV5(ax), as well as a NV center with opposite orientation in the fourth double layer, denoted by VN4(ax). NV2(ba) denotes a NV in basal orientation in the second double layer.
\label{fig:defect_models}}
\end{figure}

In this study the two most prominent cases of diamond surfaces are  considered, namely the (001) and (111) surface orientations. 
As in our previous work\cite{ko21} we build the atomistic supercell models from the bulk primitive cubic cell of diamond with a lattice constant of $a$= 3.567 {\AA} which agrees with the experimentally determined one of Holloway et al..\cite{hol91}  
Cubic diamond (Ramsdell notation\cite{ram47} 3C) has a layered structure with an ABC stacking sequence of carbon double layers in a [111] direction. A (111) oriented hexagonal bulk unit cell contains 6 carbon atoms. The supercell models for the (111) orientation contain $6\times 6$ hexagonal unit cells in the a-b plane with lattice vectors of length 15.13 {\AA}. The supercell dimension in [111]-direction (c axis) is  40.2 {\AA}  which includes an additional vacuum range of 10--12 {\AA}, depending on the considered surface termination. For 100\% N-termination we consider 13 double layers of carbon atoms and an additional monolayer of nitrogen on each of the two surfaces in the supercell, cf. Fig.~\ref{fig:defect_models}(a). This model then comprises 1008 atoms in total, 936 C and 72 N atoms. For 100\% H-coverage the model contains 13 double layers plus an additional terminating carbon layer with H atoms attached in positions ``on top'' of C atoms, cf. Fig.~\ref{fig:defect_models}(e). This model then contains a total of 1080 atoms, thereof 72 H atoms. For mixed (N,H) coverage, the terminating layer consists of a mixture of nitrogen atoms and C-H pairs, as illustrated in Fig.~\ref{fig:defect_models}(c).
The size of the slab model is about 26 {\AA} for 100\% N termination and 28 {\AA} for 100\% H termination. We enumerate the inner 13 carbon double layers as indicated by the black numbers on the left of Fig.~\ref{fig:defect_models}(a). For clarity, only half of the supercell model is displayed.

For the (001) orientation we use supercells with $6 \times 6$ cubic primitive units in the a-b plane. The corresponding lattice vectors have a length of 15.13  {\AA}. The supercell dimension in c direction amounts to 44.3 \AA\ including a vacuum range of about 10--12 {\AA}. The diamond supercell with 100\% N termination was constructed with 33 carbon monolayers plus an additional layer of nitrogen on each of the two sides of the slab. This gives 72 N and 1188  C atoms.
Fig.~\ref{fig:defect_models}(b) displays half of the supercell with the enumeration of the single C layers on the right side, in order to indicate and specify the various positions of NV centers. 
For 100\% hydrogen coverage the 33 central carbon monolayers are terminated by an additional carbon layer with H atoms attached on top of C atoms, cf. Fig. \ref{fig:defect_models}~(f). This supercell contains 1260 C and 72 H atoms in the supercell summing up to 1332 atoms in total. In the case of mixed (N,H) we follow the surface construction of 
Refs.-\cite{surf1,surf2} suggesting the formation of N-C-H bonds at the surface, 
cf.~Fig.~\ref{fig:defect_models}(d). Moreover, we examine homogeneous and clustered distributions of N atoms on the surface, which is discussed in detail in Sec.\ref{sec:suface_configs}. The lowest energy configurations of the mixed surfaces, being those with a homogeneous distribution of N-C-H groups, are illustrated in the inset of Fig.~\ref{fig:elect_aff}. 

Note that for the pure hydrogen or nitrogen terminations, our atomic surfaces structures correspond to those already discussed in the literature 
(eg.~see Ref.\,[\onlinecite{chou17b}] for (111) and Ref.\,[\onlinecite{sta15}] for (001)).

\begin{figure*}{}
\begin{center}
\setlength{\unitlength}{1mm}
\begin{picture}(170,60)(0,0)
\put(5,55){(a)}
\put(90,55){(b)}	
\put(5,0){\includegraphics[width=\columnwidth,draft=false]{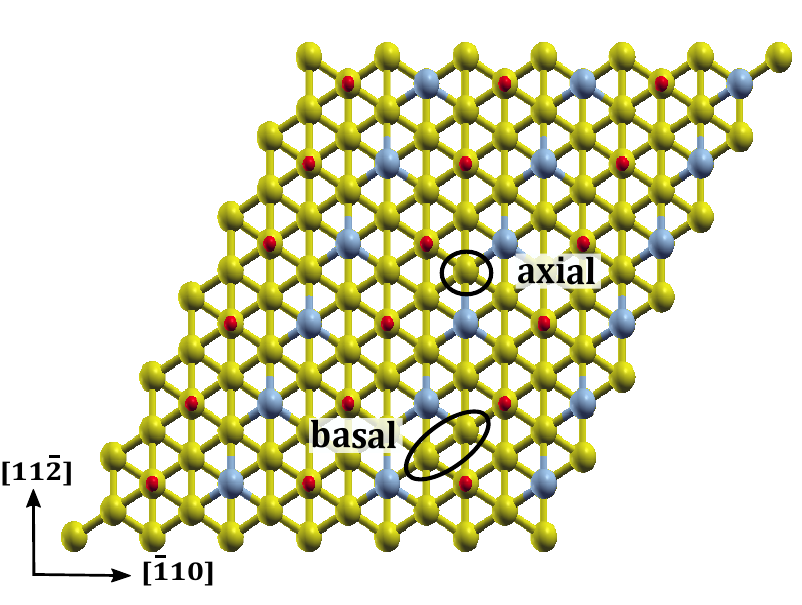}}  
\put(90,0){\includegraphics[width=\columnwidth,draft=false]{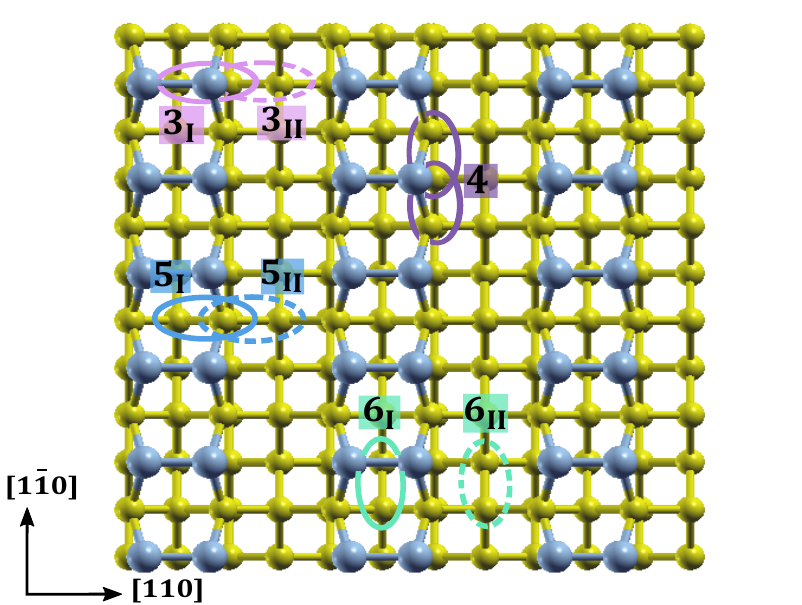}} 
\end{picture}
\end{center}
\caption{(Color online) (a) Top view on a specific (low energy) structure model of a (111) surface with a coverage of 50\% H and 50\% N. An axial and a basal orientation of a NV center are shown as a black circle and ellipse, respectively.
	(b) Top view on the atomic structure model of the (001) 
surface with 100\% nitrogen termination. The seven different possible locations of NV relative to the surface are illustrated. The possibilities shown here repeat every four monolayers. 
However, the differences between the variants become less important with growing distance from the surface.
\label{fig:defect_models2}}
\end{figure*}

In order to determine the interplay of the surface and the NV$^-$ center we considered all symmetry-inequivalent NV positions within the supercells. These positions vary in distance and orientation with respect to the surface.  
For the (111) orientation and mono-atomic termination there are four possible orientations due to the high symmetry of the surface. NV centers are either oriented parallel to the [111] direction (called \textit{axial}) or inclined by about 109.5$^\circ$ with respect to the c-axis (called \textit{basal}). In both axial and basal configurations there are two possibilities to place the NV center, with either the substitutional N atom or the C vacancy closer to the surface. These are denoted by NV and VN, respectively. 
The axial NV have the full C$_{3v}$ symmetry as in the bulk crystal while the basal NV has a reduced  C$_{1h}$ symmetry due to the presence of the surface.
In the case of mixed (H,N) terminations all the NV centers have the reduced C$_{1h}$ symmetry which implies a splitting of certain energy levels near the surfaces.

For the (001)-oriented surface with 100\% nitrogen termination the situation is more complicated. There are seven sets of NV centers which differ in their relative orientations with respect to the N-N pairs on the surface, as illustrated in Fig.~\ref{fig:defect_models2}(b). Due to the layer stacking sequence in c direction, there are equivalent orientations every four layers.  For surfaces with mixed (N,H)-termination the situation becomes even more complicated.
Of course, with increasing distance to the surface all the properties of the various NV centers assimilate to those of a NV center in the bulk crystal. 

In the following sections we will analyze the properties of NV$^-$ centers in terms of their distance to the surface. 
For this study we define the distance between a NV center and the surface by the distance of the vacancy (representing the geometrical center of the defect complex) to the first full layer of carbon atoms. 
For example, the NV center denoted VN3$_{{\rm I}}$ in Fig.~\ref{fig:defect_models}a) has a distance of 1.78 {\AA} and NV2(ba) in Fig.~\ref{fig:defect_models}b) has a distance of 4.12~{\AA}.
In our supercell models the NV centers can be placed at a maximum distance of about 14 {\AA} to the surface.
The distance between two periodic images of NV centers in the a-b plane is about $15.1$ {\AA}, i.e. the length of the respective lattice vectors of the supercells.

\subsection{Computational details}\label{sec:computation}

For the structural relaxation of the atomistic supercell models and
the calculation of the physical parameters of interest we use the Vienna Ab Initio Simulation Package 
(VASP)\cite{kr96,kr99}. The Bloch waves of the valence electrons are expanded in a plane-waves basis (with a cutoff energy of 420 eV) and the interactions of the valence electrons and the ionic cores are included by projector-augmented-waves (PAW) potentials.\cite{bl94} 
The exchange-correlation energy and potential are treated 
in the generalized gradient approximation (GGA), as given by Perdew, Burke and Ernzerhof (PBE).\cite{Perdew1996}       

For all the supercells the Brillouin-zone integrals are evaluated only at the $\Gamma$-point with a Gaussian broadening of the energy levels by 0.05 eV. The positions of the atoms in the constant supercell volume were relaxed until the residual forces acting on them were less than 0.03 eV/\AA\ and the energy difference between two consecutive ionic relaxation steps was less than $10^{-5}$\,eV. 

The computation of the HFS tensor components $\boldsymbol{A}_{ij}^I$ and the ZFS tensor components $\boldsymbol{D}_{ij}$ was done using subroutines implemented in VASP. A summary of the key formulae can be found in the Appendix.

For the calculation of the electron affinity $\chi$ we used the pure surface-slab models without NV centers. To first order, $\chi$ is given by
\begin{equation}
\label{eqn_ea}
	\chi=E_{vac}-E_{CBM},
\end{equation}
where $E_{vac}$ is the vacuum level and $E_{CBM}$ is the energy of the conduction-band minimum (CBM).
$E_{vac}$ is determined by calculating the average electrostatic potential perpendicular to the surface and by subsequently taking the plateau value in the vacuum region (see e.g.\ Fig.\ S4 of the supplemental material of Ref.\,[\onlinecite{kav14}].)
$E_{CBM}$ is obtained from the local density of states (DOS) of carbon atoms in the respective bulk regions of our models.
Since the PBE functional underestimates the band gap of diamond by more than 1 eV, a common
practice is to add the experimental band gap value of 5.47 eV\cite{cu98} to
the calculated valence band maximum.\cite{sq06,tiw11} Theoretical results using this correction scheme are indicated by the label \emph{+corr} in Table \ref{table_ea}. Our results do not include any such post-processing correction. A detailed discussion on the calculation of the EA can be found in Ref.\,[\onlinecite{sq06}].

In order to obtain and model a charged NV$^-$ center one can either consider charged supercells that contain an additional electron or include a second substitutional nitrogen atom in the structure. 
The advantage of the latter approach is that the slab in total is charge neutral. However, the extra N atom introduces additional symmetry breaking which will affect the surfaces and more importantly the NV$^-$. In our extensive supercell models of about 1000 atoms we could not find a site that is distant enough such that the N atom is not affecting the properties of the NV$^-$.  
We have tested both approaches and obtained qualitatively very similar results for the formation energies, DOS and hyperfine constants but the transversal component E of the ZFS differs substantially since it is very sensitive to differences in the charge densities of the three next-neighbor C atoms to the vacancy of the NV$^-$. We have chosen the charged supercell approach for this study and can therefore at least describe the experimentally important case of axial NV centers at (111) surfaces adequately (see Sec. \ref{sec_E}). The drawback of this methodology is that we have to deal with a charged slab which manifests itself in the presence of small slopes in the ZFS parameters D and E as well as in the HFS constants A$_{ii}$, as function of the NV center position. However, the respective changes are smaller than 1\% across the inner slab, so that the quantities appear constant in the bulk regions (see Figs. in the following Secs.~\ref{sec:results:zfs} and \ref{sec:results:hyperfine}).
A detailed discussion of several approaches to simulate charged diamond-surface slabs can be found in Ref.\,[\onlinecite{chou17a}].

\section{Results and discussion}\label{sec:results}

\subsection{Electron affinity}\label{sec:results:ea}

\begin{table}
\begin{tabular}{l c c c r }
\hline \hline 
Termination       & (001) H    & ~(111) H~  &~(001) N & (111) N  \\
\hline 
This work, PBE    & -1.02      & -1.0       & 4.0 &3.5    \\
Experiments       & -1.30$^{a}$& -1.27$^{b}$&     &  \\
LSDA+corr$^{c}$   & -1.96      & -2.01      &     &  \\
HSE06$^{d}$       & -1.7       & -1.6       & 3.5 & 3.2 \\
\hline \hline
\end{tabular}
\caption{Comparison of electron affinities $\chi$ for (001) and (111) surfaces with either H- or N-termination. The experimental data is taken from  
$^{a}$Ref.~[\onlinecite{mai01}] and $^{b}$Ref.~[\onlinecite{cu98}], the theoretical literature data stems from $^{c}$Ref.~[\onlinecite{tiw11}] and 
$^{d}$Ref.~[\onlinecite{chou17a}].
\label{table_ea}}
\end{table}

\begin{figure}[]
\centerline{\includegraphics{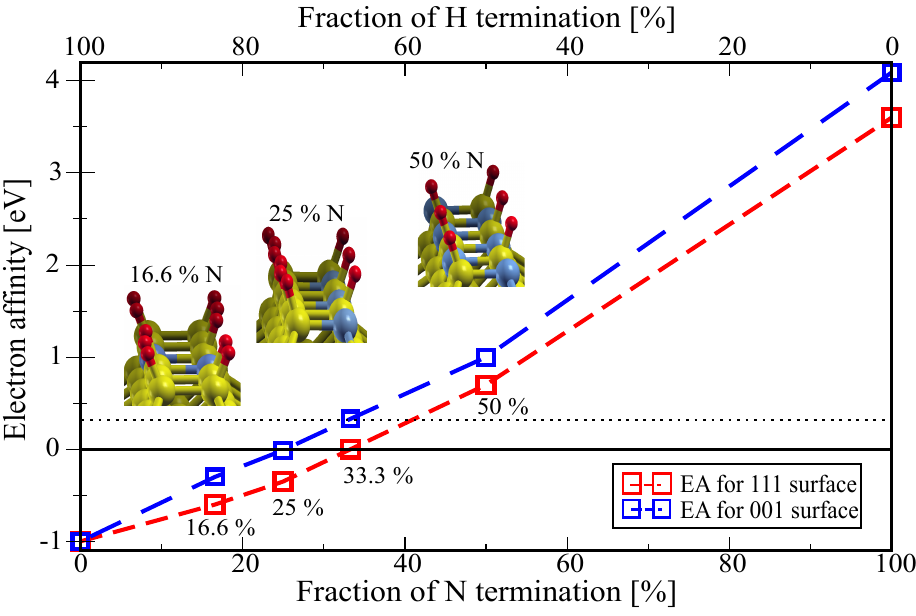}} 
\caption{(Color online) Electron affinity for (001) and (111) surfaces with mixed (N,H) coverage as function of nitrogen (hydrogen) proportion.  The three insets exemplary show the considered (001) surface terminations. The dotted line at 0.3~eV marks the average offset of our results compared to experimental values obtained for pure H  coverage, cf.~Tab.~\ref{table_ea}.
\label{fig:elect_aff}}
\end{figure}

We evaluated the EA for mixed (N,H)-coverage as function of the partial nitrogen concentration (or the N:H ratio) of the surface termination using the supercell models without any added point-defect complex. Pure hydrogen coverage leads to a negative EA and pure nitrogen coverage to a positive EA (see results in Table \ref{table_ea}).  Our point of interest is the N:H ratio at which the EA changes sign.  
Therefore, we have investigated NV$^-$ centers in slab models with 16.7\%, 20\%, 25\%, 50\% and 100\% nitrogen on the surface.
Figure \ref{fig:elect_aff} displays our results for the two considered surface orientations (001) and (111). The sign change occurs at approx.\ 33\% nitrogen for the (111) surface and at approx. 25\% for the (001) surface.

This critical value for the N:H ratio was also observed in our subsequent simulation of the NV$^-$.
For slab models with less than 25\% nitrogen on the surface the NV$^-$ defect complex was not stable even at the maximum distance of about 14 {\AA} from the surface.  The NV center could not bind the extra electron and, instead, the negative charge accumulates in the vacuum region of the supercell. However, this is not a real physical effect but an artefact of the periodic supercells which do not allow for a complete withdrawal of the charge.
For 25\% nitrogen or more we obtained self-consistent solutions with the extra electron bound to the NV defect complex. 

Chou et al.\cite{chou17a,chou17b} calculated the EA using the HSE hybrid functional and reported the sign change of the EA at 45\% nitrogen and 55\% hydrogen on the (111) surface. Their results for mono-atomic coverage are included in Tab.\ \ref{table_ea} for comparison. Our EA results for pure H coverage differ from the experimental values only by an offset of approximately 0.3 eV. Shifting our results (that were obtained using the PBE functional) by 0.3 eV yields the sign change of the EA at approx. 40\% nitrogen for the (111) surface and at approx. 33\% for the (001) surface, as illustrated by the dotted black line in Fig.\ \ref{fig:elect_aff}.

Experimental evidence for shallow stable NV$^-$ centers for slightly negative EA were recently reported.\cite{zhu16}
The experimental results for the EA are in approximate quantitative agreement with the results of our simulations which include the NV center and an extra electron in the supercell.  
The discussion in the following sections is restricted to details of \textit{stable} NV$^-$ centers near surfaces with N:H ratios of 1:3, 1:1 and 1:0, corresponding to 25\%, 50\% and 100\% nitrogen, respectively.

\subsection{Formation energies of NV$^-$ centers}\label{sec:results:energy}

\begin{figure}[]
\centerline{\includegraphics{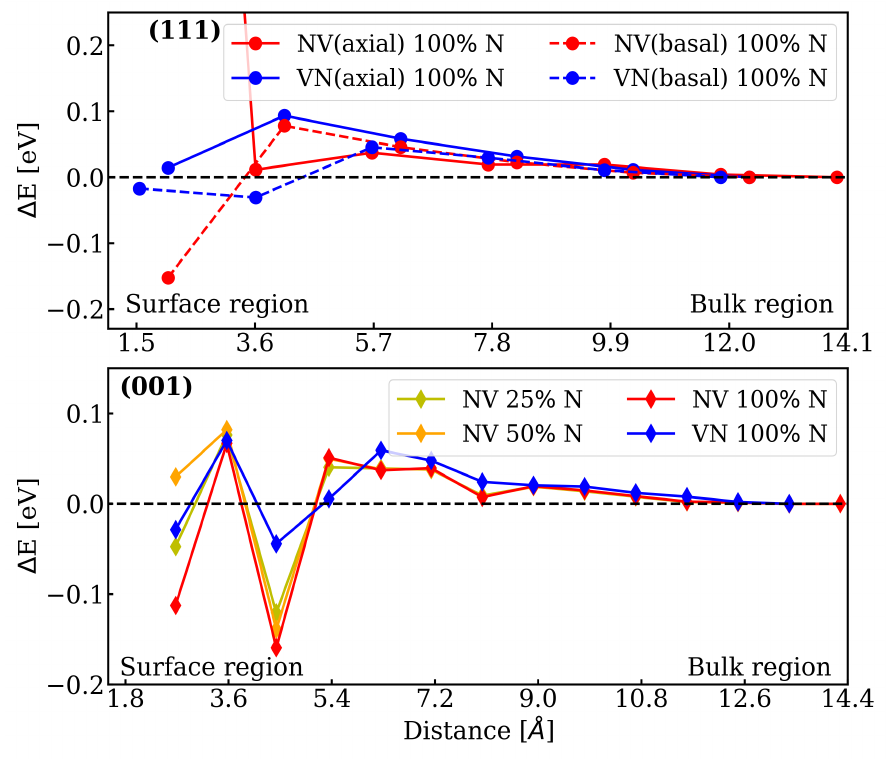}} 
\caption{(Color online) Formation energies of NV defect complexes with different orientations in the vicinity of (N,H)-terminated surfaces relative to their formation energies inside diamond at 14 {\AA} distance to the surfaces. Top: (111)-oriented surface with data points starting from double layer 1; Bottom: (001)-oriented surface with data points starting from single layer 3. For positions closer to the surfaces no stable NV center-like defect complexes were obtained.
\label{fig:form_energies}}
\end{figure}

\begin{figure}[]
\centerline{\includegraphics{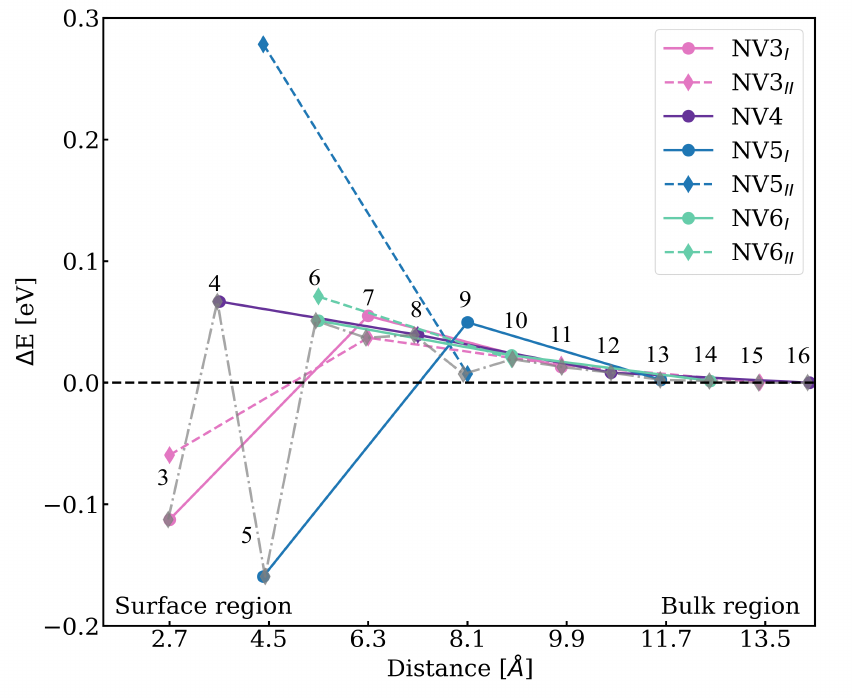}} 
\caption{(Color online) Formation energies of NV centers in the vicinity of 100\% N-terminated (001) surface. The data is divided into sets according to the different orientation with respect to the N atoms on the surface, cf.~Fig.~\ref{fig:defect_models2}(b).
The gray dashed-dotted line connects the lowest energies of the sets as function of distance (which yields the red line in the lower panel of Fig.~\ref{fig:form_energies}).
\label{fig:form_energies_groups}}
\end{figure}

We investigated the influence of the proximity of the surface on the energetic stability of the NV center by a comparison of total energies.
Figure \ref{fig:form_energies} displays the formation energies of NV$^-$ centers as function of their distance and orientation with respect to the surface. For every distance considered, only the value of the lowest-energy orientation is plotted.
Values are given relative to the formation energy of NV$^-$ at the innermost bulklike layers. Here, the double layer $n= 7$ is the reference for the (111) surface, while the monolayer $n= 16$ is the reference for the (001) surface (cf.\ Fig.\ \ref{fig:defect_models}).

NV$^-$ centers that are incorporated closer than approx. 8 {\AA} to the surface partially vary significantly in their formation energies by more than $\Delta E=\pm 0.1$ eV. Close proximity to the (111) surface causes a strong surface reconstruction which may yield large changes in energy.
For example, NV1(ax) has  $\Delta E$ $\approx$ 2.5 eV. The typical geometric arrangement of the NV$^-$ defect complex is not conserved in this case, and thus it does not make sense to classify it as a NV$^-$ center. Correspondingly, for the (001) surface, we found that defects in layers 1 and 2 lead to substantial rearrangement of the surrounding atoms, so that they can no longer be classified as a NV$^-$ defect. Therefore, we only show data points for which the local geometry of a NV$^-$ complex is preserved. 

For the (001) surfaces with NV-orientation we display the data for 25\%, 50\% and 100\% nitrogen concentrations at the surface in the lower panel of Fig.\ \ref{fig:form_energies}. It can be seen that the formation energy is practically independent of the surface chemistry. The same behavior is observed for the VN-orientation as well as for the axial and basal NVs for the (111) surfaces. As all these results are very similar, we refrain from plotting them for all considered N:H ratios for the sake of clarity.

The alternating up and down in energies observed between 2 and 6 \AA\ from the surface in the case of (001)-orientation can be understood from the seven different sets of data, corresponding to the different possible orientations of the NV$^-$ with respect to the N atom pairs on the surface, as illustrated in Fig.~\ref{fig:form_energies_groups}. Due to the specific surface reconstruction in (001)-orientation NV$^-$ centers can be accommodated in different relaxation modes within the same monolayer or with respect to the monolayers above or below. For instance, at a distance of 4.5 \AA\ the NV5$_{{\rm I}}$ is much lower in energy than NV5$_{{\rm II}}$ and the NV3$_{{\rm I}}$ and  NV5$_{{\rm I}}$ are energetically more favorable than NV4 and NV6$_{{\rm I}}$ in the neighboring monolayers. However, as can be seen, the differences between the sets of data become less pronounced with growing distance from the surface. This emphasizes the short-range effect of the surface reconstruction which does not disturb stable isolated  NV$^-$ centers which are approx. 8 \AA\ away from the surface.  

We conclude that NV$^-$ centers can be stabilized very close to the (001) and (111) surfaces since neither towards the surfaces nor towards the bulk region there is a significant energy gradient. 
Especially,  the positions NV3$_{{\rm I}}$ and NV5$_{{\rm I}}$ for (001) and NV1(ba) for (111) are the most stable positions. However, very close to the surfaces the electronic levels of the NV$^-$ centers are strongly modified as will be shown in the following.

\subsection{Eletronic levels}\label{sec:results:elec}
In bulk diamond, far from any surfaces or other defects, NV$^-$ centers have characteristic sharp electronic levels lying deep within the band gap.
The single-electron levels $a_1(2)$, $e_x$ and $e_y$ are well separated from the band edges.\cite{ga08,chou17b} Near surfaces the situation changes since surface states appear. For example in the case of hydrogen termination these lead to a wide electronic band below the CBM of bulk diamond.\cite{kav14} Such bands interfere with $e_x$ and $e_y$ if the NV$^-$ center is spatially located close enough to the surface.  

Our (111) surface models with mixed termination and at least 25\% nitrogen have an open band gap like in bulk diamond and deep lying NV$^-$ defect states that are neither close to the conduction band nor the valence band. This was already reported for pure nitrogen termination by Chou et al.\cite{chou17b} for the (111) surface.  For our (001) slab models, surface states appear at approx. 0.8 eV below the CBM for mixed termination with 25\% nitrogen. Their position changes  to 0.3 eV below the CBM for 50\% nitrogen. 

\begin{figure}[]
\begin{center}
\setlength{\unitlength}{1mm}
\begin{picture}(85,56)(0,0)
\put( 0,53){(a)}
\put(55,53){(b)}	
\put(0,0){\includegraphics[width=\columnwidth,draft=false]{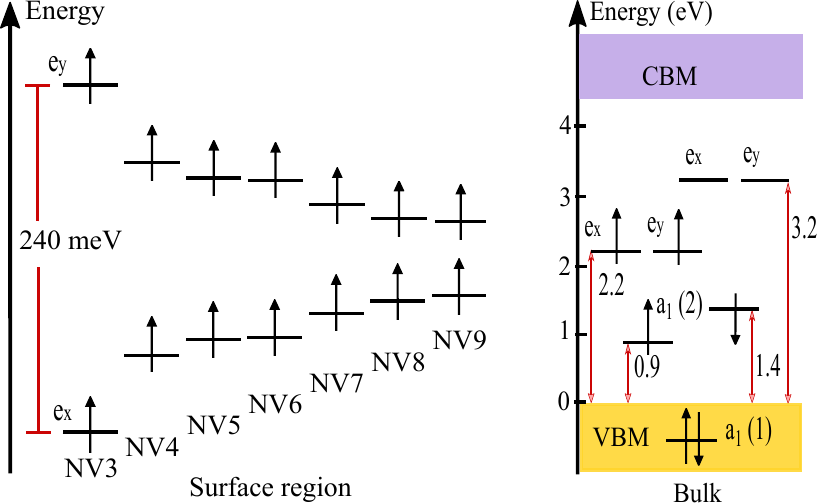}} 
\end{picture}
\end{center}
\caption{(Color online) (a) Splitting of the single-electron levels $e_x$ and $e_y$ for NV$^-$ centers near the (001) surface with 100\% N-	termination. The single-electron levels $a_1(2)$ are only marginally affected by the vicinity of the surface and therefore not shown. 
(b) Level structure of the NV$^-$ with values taken from Ref.~[\onlinecite{ga08}]. NV$^-$ energies are given with respect to the VBM.
\label{fig:split_exey}}
\end{figure}

Despite the open band gap, a NV$^-$ center that is close ($< 7$ \AA) to a surface causes structural relaxations that alter its electronic levels. In the bulk interior, the $e_x$ and $e_y$ levels are degenerate since they are formed by the equivalent three carbon atoms next to the vacancy.\cite{ga08}
This degeneracy of $e_x$ and $e_y$ is lifted near the surfaces due to the reduced symmetry. 
Figure~\ref{fig:split_exey} shows this energy splitting of the single-electron levels $e_x$ and $e_y$ for NV$^-$ centers near the (001) surface with 100\% N-termination. A maximal splitting of 250 meV of the single-electron levels $e_x$ and $e_y$ was obtained at position NV3  decreasing to 50 meV at position NV9.
For NV$^-$ centers at even larger distance to the surface, we observe that all of them also show this minimal level splitting of $\approx50$ meV as a consequence of using a \textit{finite} slab model geometry for the simulation.
NV$^-$ centers with basal orientation near a (111) surface behave similarly like NV$^-$ centers near the (001) surface. By contrast, the threefold symmetry is retained for NV$^-$ centers with axial orientation near a perfectly flat infinite (111) surface with pure N termination. 

We refrain from giving absolute values for the electronic levels not only because it is well known that the PBE functional is already underestimating the band gap of bulk diamond by more than 1 eV but also because there are further problems. In finite slab models of a few nm size there are confinement effects, self interaction effects in lateral direction, and also the presence of two polarized surfaces due to NV$^-$ in the model structure. A thorough discussion of these effects for the NV center in diamond was presented by Chou et al.\cite{chou17a}
In Fig. 5 of their paper the authors also present the defect levels of NV$^-$ in the vicinity of various surfaces which agree qualitatively with our results. 
 
\subsection{Zero field splitting}\label{sec:results:zfs}

\subsubsection{Longitudinal ZFS component D} 
The zero-field splitting component $D$ of the singlet and triplet states of the $^3A_2$  ground state of a NV$^-$ center located in a bulk diamond crystal was experimentally measured as $D$=$\frac{3}{2}D_{zz}$= 2.872(2) GHz with high precision by Felton et. al.\cite{fel09}.
Our theoretical values for bulk diamond exceed the measured value by about 3 to 6\% depending on the supercell size and other computational details. A discussion of the limitations and approximations can be found in our previous work\cite{ko21} and in Ref.~[\onlinecite{bik20}] which addresses the central problem of the so-called spin contamination.

\begin{figure}[]
\centerline{\includegraphics{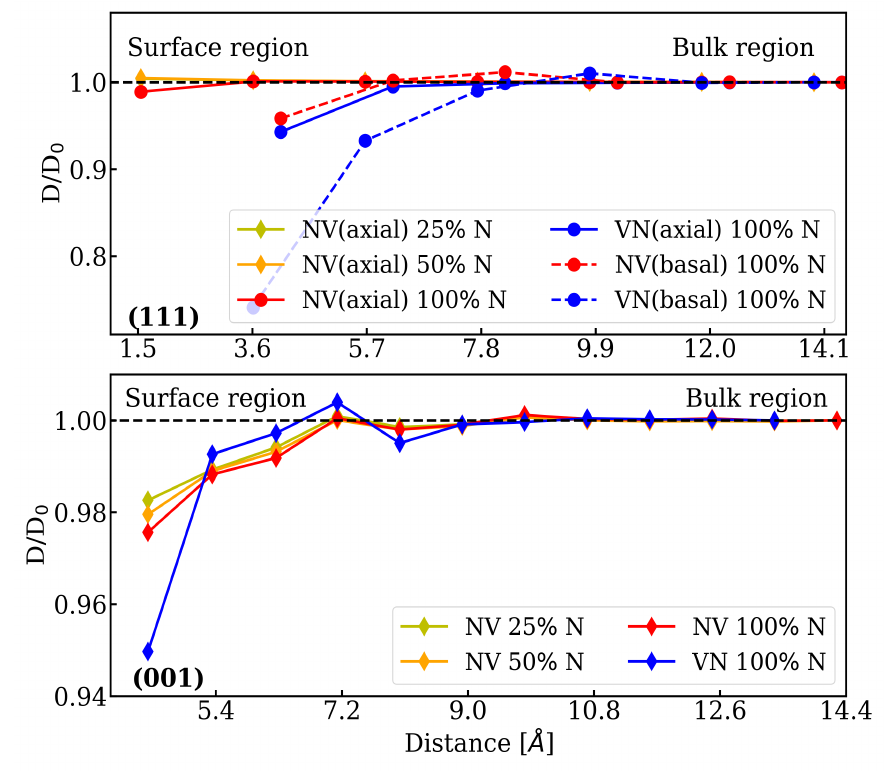}} 
\caption{(Color online) Longitudinal component $D$ of the ZFS of a NV$^-$ center for different positions relative to the surface. Top panel: (111)-orientation; Bottom panel: (001)-orientation. 
\label{fig:Dzz_zfs}}
\end{figure}

In Fig.~\ref{fig:Dzz_zfs} the axial component $D$ of the ZFS for our surface models normalized to the value $D_0$ obtained from the respective bulk supercells is shown. 
Near the surface the $D$ values are reduced, where the reduction is in all cases more pronounced for the VN-oriented NV$^-$ centers than for the NV-oriented ones. This can be understood from the fact that the main contribution to the ZFS originates from the three C atoms next to the vacancy. Each of them has a magnetic moment of about 0.52 $\mu_B$, amounting together to 78\% of 2 $\mu_B$.
For the NV$^-$ centers with NV-orientation these three C atoms lie further away from the surface and are thus situated in a more bulk-like neighborhood than the ones of the VN-orientation. 

The influence of the surface on the axial component $D$ of the ZFS is limited to about 8 \AA\ below the surface irrespective of the surface orientation and its chemical termination.
As shown in Fig.~\ref{fig:Dzz_zfs}, a very weak dependence on the N:H ratio is found for both (001) and  (111) surfaces.
 
\subsubsection{Transversal ZFS component E}\label{sec_E}

In perfect diamond and zero external magnetic field $B_{{\rm ext}}$ along the NV$^-$ axis the $m_{{\rm S}}=\pm1$ states are degenerate, where $m_S$ denotes the spin projection on the NV$^-$ axis. In experiments a magnetic field can be sensed directly by measuring the Zeeman-splitting (namely by a splitting of 5.6 MHz per 1 Gauss field) of the $m_{{\rm S}}=\pm1$ states.\cite{lee19}
However, crystal imperfections may lead to finite values of the transversal component $E$ and thus additional splittings as discussed e.g. in our previous work.\cite{ko21} 
The total splitting of the energy levels for the spin $S=1$ system reads
\begin{eqnarray}
	E_{m_{{\rm S}}=\pm1}&=&D\pm\sqrt{E^2+(g \mu_B B_{{\rm ext}})^2}
\end{eqnarray} 
and $E_{{\rm m_S=0}}=0$.

\begin{figure}[]
\centerline{\includegraphics[width=\columnwidth]{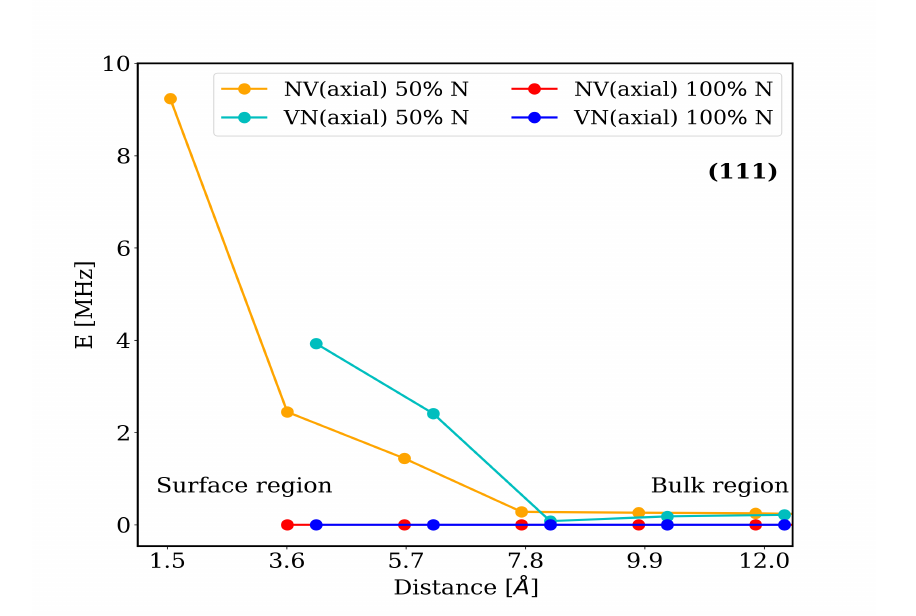}} 
\caption{(Color online) Transversal component E of the ZFS of a NV$^-$ center with axial orientation near (111)-oriented surface.
\label{fig:E_zfs}}
\end{figure}

The calculation of $E$ is computationally demanding. 
It does not only suffer from systematic errors like the aforementioned spin contamination\cite{bik20} but in our situation the charged NV$^-$ defect induces dipole charges at the two surfaces of the slab models.
Theses dipoles cause an artificial electric field across the slab.
Only for axial NV$^-$ centers the effect of the resulting dipole field cancels due to symmetry, since all three C atoms near the vacancy, which are responsible for the main contribution to $E$, experience the same field.

Therefore, we present the calculated values of the transversal component $E$ of the ZFS for the $^3A_2$ ground state only for NV$^-$ centers with axial orientation. In Fig.\ \ref{fig:E_zfs} one can see that $E$ depends crucially on the specific (111) termination. For 100\% N-termination the threefold symmetry is conserved and thus $E=0$ MHz, independent of the distance to the surface. For mixed (N,H)-terminations this symmetry is broken which leads to small splittings on the order of 10 MHz close to the surface. However, these values quickly decay to 0 MHz at larger distances to the (111) surface.

\subsection{Hyperfine splitting connected to $^{14}{\rm N}$ and  $^{13}{\rm C}$ }\label{sec:results:hyperfine} 

\begin{figure}[]
\centerline{\includegraphics[width=\columnwidth]{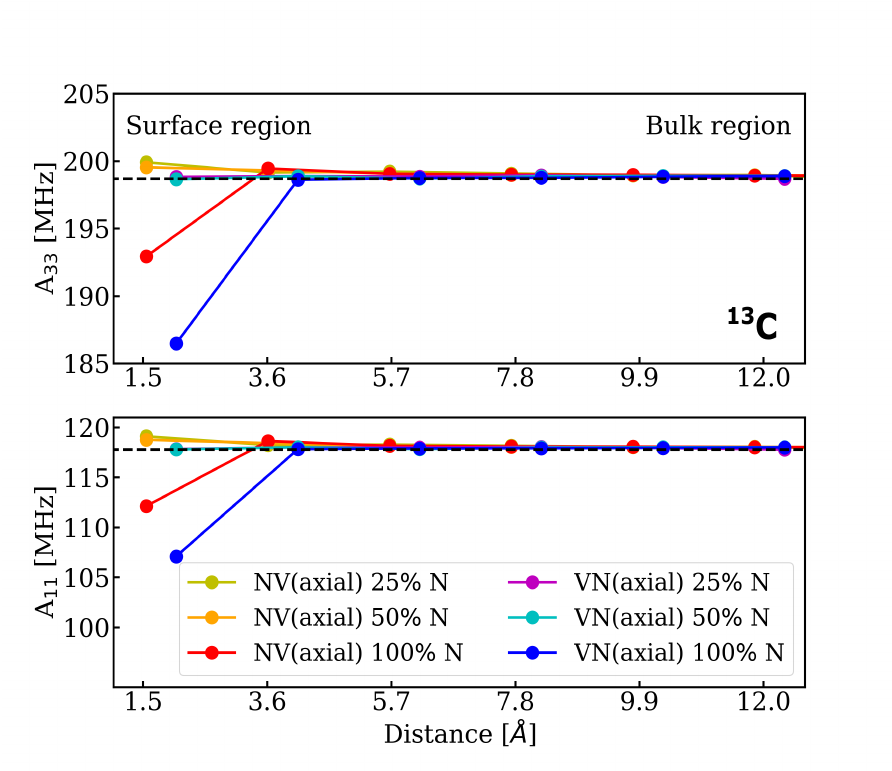}}  
\caption{(Color online) Hyperfine structure constants $A_{ii}$ for $^{13}$C for axially oriented NV$^-$ centers near the (111) surfaces with different termination.      
\label{fig:Aij_C13}}
\end{figure}

Calculated hyperfine structure constants for the NV$^-$ centers in the cubic diamond crystal are in good accordance with recent experimental work of Felton et al.\cite{fel09}, as discussed in our previous work\cite{ko21}. 
This is a good starting point for the study of the influence of the (111) and (001) surfaces on the HFS.
Among the NV-center properties investigated in this work, the HFS constants are the least dependent on the proximity to the surfaces. As an example, the HFS related to $^{13}$C for the practically most relevant case of axial NV centers near (111) surfaces is shown Fig.~\ref{fig:Aij_C13}. For distances of approx.~4 {\AA} or more the NV$^-$ centers take their typical bulk values. For NV$^-$ centers closer to the surface we observe  that the A$_{ii}$ values are more affected for the VN-oriented ones. This is conceivable since for VN-orientation the C atoms are closer to the surface than in the NV-orientation. Basally oriented NV centers show modified HFS for distances of approx.~6 {\AA}, so one double layer deeper into the bulk.

The results for the Hyperfine splitting connected to $^{14}{\rm N}$ are very similar to the ones of $^{13}{\rm C}$, which means that the change of the $A_{ii}$ is restricted to the first 6 {\AA}.
However, since the absolute frequencies ($A_{11}$= -2.16 MHz and $A_{33}$= -1.73 MHz\cite{ko21}) are two orders of magnitude smaller compared to the  $A_{ii}$ of $^{13}$C
the relative increase is in few case of order 20\% very close to the surfaces ($A_{33}$= -1.35 MHz for VN2(ax) near 100\% nitrogen (111) surface).  

We obtain a similar result for the (001) surface. Here the deviation of A$_{ii}$ is ~ 5\% for $^{13}{\rm C}$ and ~ 10\% for $^{14}{\rm N}$ for the NV$^-$ centers closer than 5 {\AA} to the surface, regardless of the surface termination. In conclusion, we observe that the HFS constants converge quickly to their bulk value with increasing distances to the surface.

\subsection{Limitations of our model}

The results presented above where obtained within the framework of DFT (PBE-GGA) calculations and do not take into account some important dynamical effects which are encountered in real diamond NV-center systems. In order to achieve a high magnetic sensitivity,  the coherence time (T$_2$) of the NV center needs to be long enough (in the range of $\mu s$). An important decoherence mechanism which affects T$_2$ is caused by fluctuating unpaired surface spins due to impurities\cite{ban72} and not atomically flat surfaces.\cite{osi09}
Romach et al.\cite{rom15} studied the surface induced noise and associated the high frequency part to phonon effects because this is significantly suppressed at 10 K. They assigned the temperature independent low frequency part of the noise to a surface spin bath. 
Since in our static DFT calculations the atomic nuclei are kept at rest ($T=0$ K) and we restrict our investigations to atomically flat and defect-free surfaces near the NV centers of interest, the dynamical effects mentioned above are excluded. 
Nevertheless, our results can be considered as an ideally static limit: a perfectly flat N-doped surface does not have unpaired spins and thus avoids dynamical surface disturbances on NV centers.

\section{Summary}\label{sec:summary}

We have studied the influence of the proximity of diamond (111) and (001) surfaces with various chemical (N,H)-terminations on the formation energies, densities of states, zero-field splittings and hyperfine structure constants of NV$^-$ centers in diamond. For surfaces with less than approx.\ 25\% nitrogen the simulations did not yield stable negatively charged NV$^-$ centers but neutral ones instead. 

The analysis of formation energies shoes that NV$^-$ centers may be stable at positions a few {\AA} below the (001) and (111) surfaces. There is no steep energy gradient, meaning the energy differences with respect to the bulk region are very small. The result for the formation energies is that formation of NV$^-$ centers closer than 6 {\AA}  to the surface occurs at a maximum energy gain of 0.15 eV. On the other hand, for distances larger than $\sim$ 8 {\AA}  the NV$^-$ formation varies only in the range of 20 meV with respect to the bulk region.

Moreover, very close to the surfaces the electronic levels of NV$^-$ centers get modified. 
The most prominent feature in the electronic structure of NV centers is the splitting of the single-electron levels $e_x$ and $e_y$. 

Furthermore, the HFS parameters $A_{ii}$ connected to  $^{13}C$ (and $^{14}N$) result in a typical decrease (increase) of 5\% to 20\% relative to the bulk values. 
These quantities are influenced by the proximity of the (001) and (111) surfaces only in a very short range of approx.\ $4$ {\AA}. 

The ZFS parameters are affected by the surfaces on a longer range. The axial component $D$, which is indicative for the singlet-triplet splitting, can be reduced up to approx. 25\% near the surfaces. 
For axially oriented NV centers at (111) surfaces the transversal component $E$ of the ZFS is small (few MHz).

Finally, the dependence of all considered quantities on the surface orientation (001) or (111) and its specific chemistry, namely the N:H ratio of the surface termination, is weak.

In conclusion, we believe that our analysis improves the understanding of very shallow NV centers in diamond and the data for the DOS, ZFS and the HFS may help to interpret experimentally obtained spectra. 
Our results show that axial NV centers near the flat 100\% N-terminated (111) surface are the ideal choice for NV-based quantum sensing applications since they feel the least influence by the proximity of the surface due to their high symmetry.

\section{Acknowledgments}
Financial support for this work was provided by the Fraunhofer Lighthouse Project Quantum Magnetometry (QMag).
The calculations were partially performed on the supercomputer ForHLR funded by the Ministry of Science, Research and the Arts Baden-W\"urttemberg and by the Federal Ministry of Education and Research.

\section{Appendix}\label{sec:app}
In this appendix we shortly discuss the influence of different surface configurations at fixed N:H ratio and summarize the notation and calculation of the HFS and the ZFS.
For more detailed information on the HFS and ZFS in diamond see e.g. Refs.\,[\onlinecite{lou78,iva14,ko21}].

\subsection{Properties of different (N,H)-configurations at the (001) surface}\label{sec:suface_configs}

\begin{figure}{}
\begin{center}
\setlength{\unitlength}{1mm}
\begin{picture}(85,135)(0,0)
\put(0.1,0){\includegraphics[width=\columnwidth]{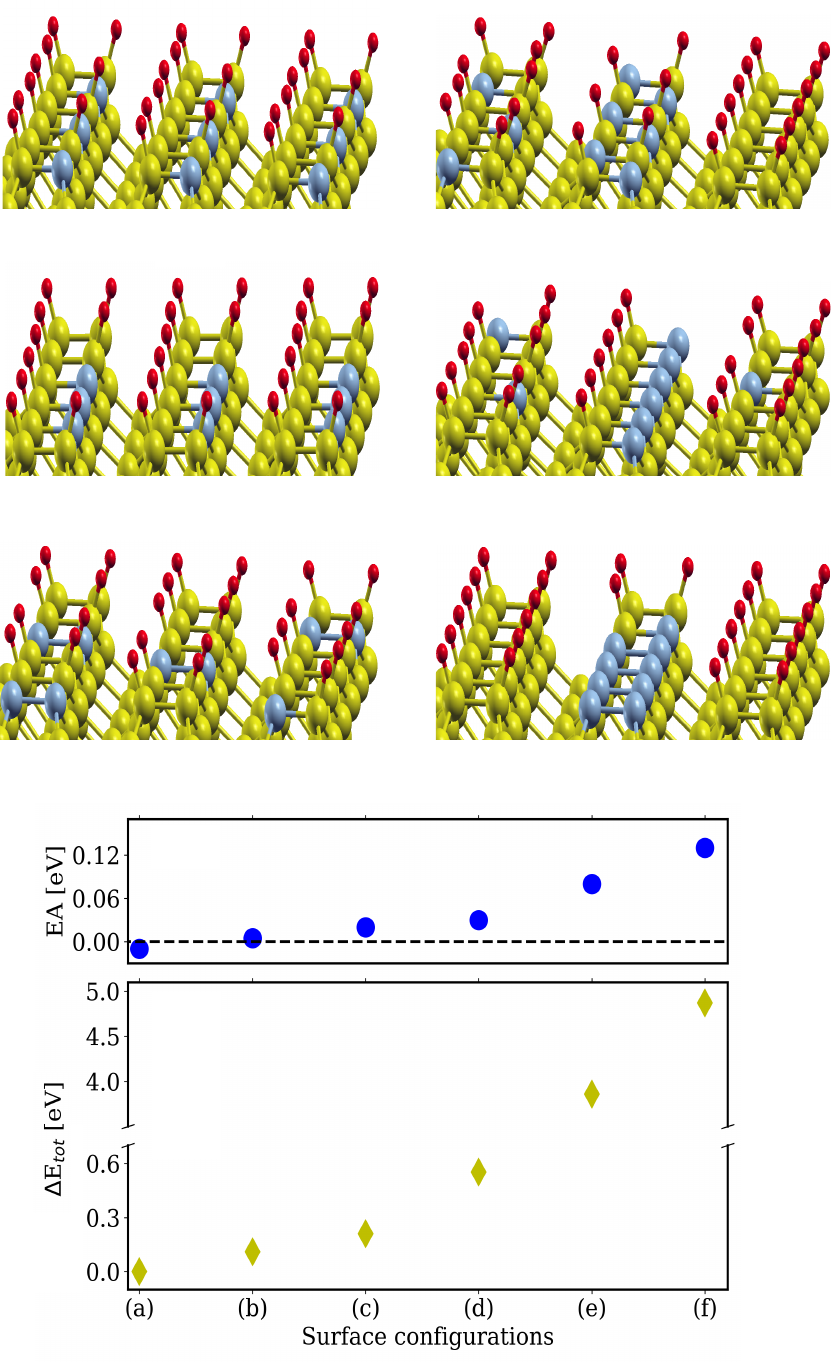}}
\put( 16.5,114){(a)}
\put(60.5,114){(b)}	
\put( 16.5,87){(c)}
\put(60.5,87){(d)}	
\put( 16.5,60){(e)}
\put(60.5,60){(f)}	
\put( 0.001,55){(g)}
\put(0.001,38){(h)}	
\end{picture}
\end{center}
\caption{(Color online)  (a)-(f) Different configurations of N and H atoms on the (001) surface with a fraction of 25\% N. (g) Electron affinity (EA) and (h) energy difference relative to configuration (a) which has the lowest-energy. Configuration (a) was chosen for the study described in the main text. 
\label{fig:surf_conf}}
\end{figure}
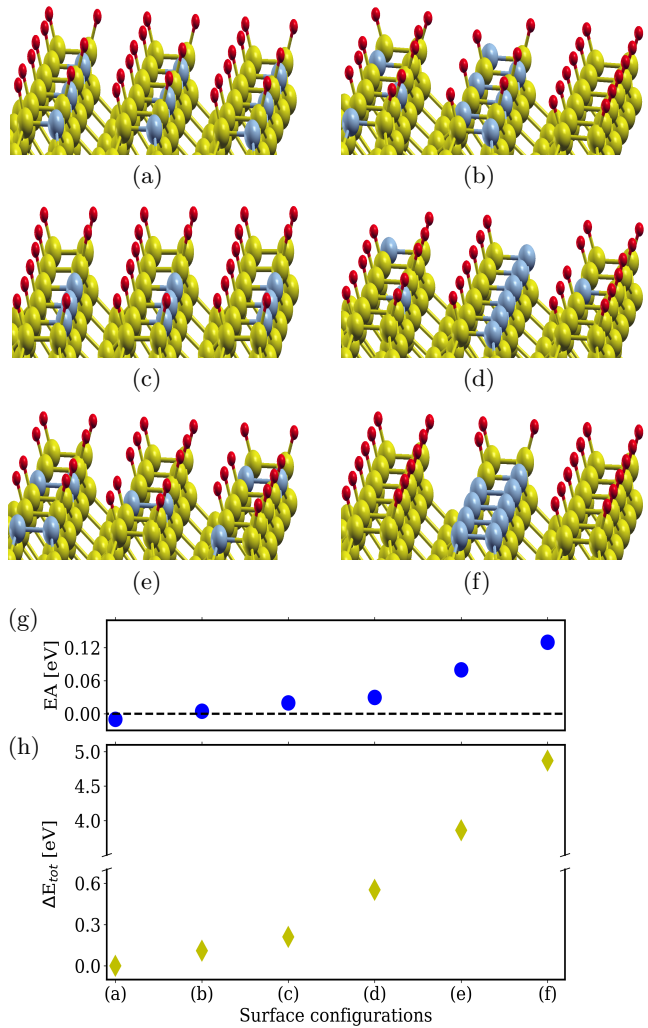

In order to obtain the optimal surface configuration for cases of mixed (N,H)-coverage we examined different orderings of N and H atoms on the (001) surface with a fraction of 25\% N atoms. As a starting point we considered surfaces with homogeneously distributed N-C-H bonds and subsequently increased the clustering of N atoms as depicted in Fig.~\ref{fig:surf_conf}(a)-(d). In addition, the formation of N-N bonds as shown in Figs.~\ref{fig:surf_conf}(e) and (f) was tested. A general trend in all these configurations is that with increasing clustering of N atoms on the surface the formation energy increases. In the case of configurations with N-N bonding the energy difference increases even up to 5~eV. Therefore, the N-C-H bonding on the (001) surface is energetically preferred and was chosen for the further calculations in our study (see (a) of Fig.~\ref{fig:surf_conf}). 

Moreover, although the change of the EA for the tested surface configurations is small, one can see that whenever the clustering of N atoms is increased (cf.\ Figs.\ref{fig:surf_conf}(a)-(d)) also the electron affinity of the surface increases. Note, that the arrangements (e) and (f) have direct N-N bonds which increase the EA even more. In that sense, a purely N terminated surface can be regarded as a huge cluster of N atoms and it is conceivable that such a configuration has a positive electron affinity (see Fig.~\ref{fig:elect_aff}). 

Throughout the paper we have used the energetically favorable highly ordered (N,H)-configurations for the (001) and for the (111) surface slab cells with NV defect. Note that already at a few carbon layers below the surface the individual (N,H)-configuration becomes unimportant.

\subsection{Hyperfine interaction}\label{sec:hyperfine}
The HFS tensor $\boldsymbol{A}^I$ describes the interaction between a nuclear spin S$_I$ 
and the electronic spin distribution. The hyperfine interaction between 
a nuclear spin S$_I$ (i.e. the nuclear spin of $^{13}$C or $^{14}$N)
and the electronic spin distribution S$_e$ (here NV$^-$ defect state) can be modeled with the Hamiltonian $H_{{\rm HFS}}= S_e  \boldsymbol{A}^I  S_I$.
The hyperfine structure tensor components A$_{ij}^I$ for a nucleus $I$ are
\begin{equation}\label{eqn_HF}
A_{ij}^{I}=\frac{\mu_0 \gamma_{I} \gamma_e}{2S}\int d^3r ~ n_{S}({\bm r}) \left[ \left(\frac{8\pi \delta(r)}{3}\right) +\left( \frac{3x_ix_j}{r^5}-\frac{\delta_{ij}}{r^3}\right) \right],
\end{equation}
where the first term in brackets is the Fermi-contact term and the second bracket is
the magnetic dipole-dipole term.
Here, $n_S$ denotes the spin density associated with spin state S,
$\mu_0$ is the vacuum magnetic permeability, $\gamma_e$ the gyromagnetic ratio of 
the electron and $\gamma_{I}$ the gyromagnetic ratio of the nucleus. In this work the values $\gamma{(^{13}{\rm C})}/2\pi$  = 10.7084 Mhz/T and 
$\gamma{(^{14}{\rm N})}/2\pi$  = 3.077 Mhz/T are used.\cite{ber04}

\subsection{Zero-field-splitting}\label{sec:zfs}
The ZFS describes the loss of degeneracy of the
electronic levels of the spin triplet state of the NV$^-$ (with different values of magnetic spin quantum number m$_{\rm S}$= 0,+1,-1) in the absence of an external magnetic field. 
It arises from the presence of unpaired electrons and can be modeled by the Hamiltonian
\begin{equation}\label{zfs}
H_{{\rm ZFS}}= \frac{\mu_0 g^2 \mu_B^2}{4\pi r^5}  \bigl[3( S_1\cdot r)(S_2\cdot r)-(S_1\cdot S_2)r^2\bigr],
\end{equation}
which describes the effect of electron-electron repulsion by magnetic dipole-dipole interaction.
Here, $r=r_1-r_2$ is the spatial distance between the spins, $S_i =\frac{1}{2}[\sigma_x, \sigma_y, \sigma_z]$ is the spin operator
vector of particle i, $\sigma_j$ (j=x, y, z) are the Pauli matrices, and g is the Land\'e factor.
One can separate the spatial and spin dependencies in equation (\ref{zfs}) and write the Hamiltonian in the form $H_{{\rm ZFS}}= S  \boldsymbol{D} S$,
where $S=S_1+S_2$ is the total spin and $\boldsymbol{D}$ describes the dipolar spin-spin interaction.
The tensor $\boldsymbol{D}$ is symmetric and traceless and thus can be diagonalized. 
 
In general, one can split $H_{{\rm ZFS}}$ into a longitudinal component of the magnetic dipole-dipole interaction (named $D$) and a transversal component (named $E$):
\begin{equation}
D= \frac{3}{2} D_{zz} ~~ {\rm and} ~~  E=  \frac{1}{2}(D_{xx}-D_{yy}),
\end{equation}
where the $D_{ii}$ are the diagonal elements of the tensor $\boldsymbol{D}$.

\end{document}